\documentclass[%
print,
 amsmath,amssymb,
 aps,
]{revtex4}

\usepackage{graphicx}% Include figure files
\usepackage{dcolumn}% Align table columns on decimal point
\usepackage{bm}% bold math
\usepackage[landscape,papersize={297.1mm,210mm},left=1.9cm,right=1.6cm,top=2.7cm,bottom=2.8cm]{geometry}
\usepackage[                   %dvipdfm, pdflatex,pdftex这里决定运行文件的方式不同
            pdfstartview=FitH,
            colorlinks, %注释掉此项则交叉引用为彩色边框(将colorlinks和pdfborder同时注释掉)
            pdfborder=001,   %注释掉此项则交叉引用为彩色边框
            linkcolor=blue,
            anchorcolor=blue,
            citecolor=blue,
            urlcolor=blue
            ]{hyperref}

\begin{document}
\title{Lower bounds of concurrence for $N$-qubit systems and the detection of $k$-nonseparability of multipartite quantum systems}

\author{}
\author{Xianfei Qi, Ting Gao}
\email{gaoting@hebtu.edu.cn}
\affiliation {College of Mathematics and Information Science, Hebei
Normal University, Shijiazhuang 050024, China}
\author{Fengli Yan}
\email{flyan@hebtu.edu.cn}
\affiliation {College of Physics Science and Information Engineering, Hebei
Normal University, Shijiazhuang 050024, China}

\begin{abstract}
Concurrence, as one of entanglement measures, is a useful tool to characterize quantum entanglement in various quantum systems. However, the computation of the concurrence involves difficult optimizations and only for the case of two qubits an exact formula was found. We investigate the concurrence of four-qubit quantum states and derive analytical lower bound of concurrence using the multiqubit monogamy inequality. It is shown that this lower bound is able to improve the existing bounds. This approach can be generalized to arbitrary qubit systems. We present an exact formula of concurrence for some mixed quantum states.  For even-qubit states, we derive an improved lower bound of concurrence using a monogamy equality for qubit systems. At the same time, we show that a multipartite state is $k$-nonseparable if the multipartite concurrence is larger than a constant related to the value of $k$, the qudit number and the dimension of the subsystems. Our results can be applied to detect the multipartite $k$-nonseparable states.
\end{abstract}

\pacs{ 03.67.Mn, 03.65.Ud, 03.67.-a}

\maketitle

\section{Introduction}
Quantum entanglement represents one of the  most striking features of quantum systems that has no classical counterpart. During the past two decades, researchers have paid concentrated effort on it and made impressive progress \cite{QCI.2010,RMP81.865,PR474.1,JPA47.424005}. An important aspect of the research on quantum entanglement is entanglement detection and quantification. Concurrence is one of the most important entanglement measures. It was introduced by Wootters \cite{PRL78.5022} who established concurrence as an entanglement measure for two-qubit states. Later, generalization to the bipartite higher-dimensional systems \cite{PRA64.042315} was proposed. A landmark result in the theory of entanglement measures was the derivation of an elegant formula for the concurrence of an arbitrary mixed states in a system of two qubits by Wootters \cite{PRL80.2245}. For arbitrary high-dimensional mixed states, analytical formulas of concurrence are available only for some highly symmetric states \cite{PRL85.2625,PRA64.062307,PRA67.012307} due to the extremizations involved in the computation.

Because the exact formulas of concurrence are very difficult to be found, it is of great significance to get the lower bounds of concurrence. An analytic lower bound for the concurrence of mixed bipartite quantum states in arbitrary dimensions was derived \cite{PRL92.167902} and this bound can be  tightened by numerical optimization over some parameters. In \cite{PRL95.040504} authors derived an analytical lower bound for the concurrence of bipartite quantum states in arbitrary dimension and established a functional relation relating concurrence, the Peres-Horodecki criterion, and the realignment criterion. A lower bound of concurrence based on local uncertainty
relations criterion was derived \cite{PRA75.052330} and was further improved \cite{PRA76.012334}. Based on  Wootters' idea, lower bounds on the concurrence for bipartite systems were presented \cite{PRA67.052308}. In \cite{PRA84.062322} the authors proposed a method to construct
a hierarchy of lower bounds of concurrence for arbitrary dimensional bipartite mixed states in terms of the concurrence
of all the lower-dimensional mixed states extracted from the given mixed states.

 The concurrence is generalized to multipartite quantum systems \cite{PRL93.230501}. Calculation of concurrence for multipartite quantum states is more difficult, and quite less has been known about the concurrence of multipartite mixed state. In \cite{PRA74.050303}, the authors derived an analytical lower bound for the concurrence for tripartite quantum mixed states.  As a generalization of the Wootters formula,  lower bounds of the concurrence
for higher dimensional systems as well as for multipartite systems were derived in \cite{PRL109.200503}. In terms of the monogamy inequality \cite{PRA61.052306,PRL96.220503} of concurrence, analytical lower bounds of concurrence of qubit systems were presented in \cite{QIP13.815}.

The separability problem for multipartite and high-dimensional systems is also important in quantum information theory. This problem is more complicated than bipartite systems owing to the complicated structure of multipartite entangled
states. For instance, one can discuss it with the notions of $k$-partite separability and $k$-nonseparability.  Although some $k$-separability criteria  have been introduced (see e.g. \cite{QIC10.829,PRA82.062113,EPJD61.765,EPL104.20007,PRA91.042313,SR5.13138,PRA93.042310}), the number of computable measure quantifying multipartite entanglement is rare. Ma \textit{et al} \cite{PRA83.062325,PRA85.062320} defined a generalized
concurrence called the genuine multipartite entanglement~(GME)~concurrence as an entanglement measure that
distinguishes genuine multipartite entanglement from partial
entanglement. Gao \textit{et al} \cite{PRA86.062323} defined $k$-ME concurrence, which
satisfies important characteristics of an entanglement measure,
such as the entanglement monotone, vanishing on $k$-separable
states, invariant under local unitary transformations,
convexity, subadditivity, and being strictly greater than zero
for all $k$-nonseparable states. Combining $k$-ME concurrence with permutation invariance, a lower bound was given on entanglement measure by using the permutation-invariance part of a state and that lower bound can apply to arbitrary multipartite systems \cite{PRL.112.180501}. At the same time, the concept of ``the permutationally invariant part of a density matrix" is proven to be more powerful because of its basis-dependent property \cite{PRL.112.180501}.

In this paper, we investigate multipartite concurrence for qubit systems and the detection of multipartite $k$-nonseparable states. We provide an analytical lower bound of concurrence for four-qubit states based on monogamy inequality. The result is generalized to multipartite case. We present an exact formula of concurrence for some mixed quantum states. For even-qubit states, we derive an improved lower bound of concurrence using a monogamy equality for qubit systems. At the same time, we show that if the multipartite concurrence is larger than a constant related to the value of $k$, the qudit number and the dimension of the subsystems, then the state must be $k$-nonseparable.

\section{Preliminaries}

Before we state the main results, an introduction of the concepts and notations that will be used in
this paper is necessary. Throughout the paper, we consider a multipartite quantum system with state space $H=H_{1} \otimes H_{2} \otimes \ldots \otimes H_{N}$, where $H_{i}~(i=1,2,\ldots,N)$ denotes $d$-dimensional Hilbert spaces. A $k$-partition $A_{1}\mid A_{2}\mid \cdots \mid A_{k}$ (of $\{1,2,\ldots,N\})$ means that the set $\{A_{1},A_{2},\ldots,A_{k}\}$ is a collection of pairwise disjoint sets, and the union of all sets in $\{A_{1},A_{2},\ldots,A_{k}\}$ is $\{1,2,\ldots,N\}$ (disjoint union $\bigcup_{i=1}^{k}A_{i}=\{1,2,\ldots,N\}$). An $N$-partite pure state is called $k$-separable if there is a $k$-partition $A_{1}\mid A_{2}\mid \cdots \mid A_{k}=j_{1}^{1}\cdots j_{n_{1}}^{1}\mid j_{1}^{2}\cdots j_{n_{2}}^{2}\mid \cdots \mid j_{1}^{k}\cdots j_{n_{k}}^{k}$ such that $$|\psi\rangle=|\psi_{1}\rangle_{A_{1}}|\psi_{2}\rangle_{A_{2}}\cdots |\psi_{k}\rangle_{A_{k}},$$
where $|\psi_{i}\rangle_{A_{i}}$ is the state of subsystems $A_{i}$ and disjoint union $\bigcup_{t=1}^{k}A_{t}=\bigcup_{t=1}^{k}\{j_{1}^{t},j_{2}^{t}\cdots j_{n_{t}}^{t}\}=\{1,2,\cdots,N\}$. An $N$-partite mixed state $\rho$ is $k$-separable if it can be written as a convex combination of $k$-separable pure states, where pure states might be $k$-separable with respect to different partitions. In general, $k$-separable mixed states are not separable with regard to any
specific partition. A $k$-nonseparable mixed state is defined to be one that cannot be written as a convex combination of $k$-separable pure states. Specially, if an $N$-partite state is not two-separable
(biseparable), then it is called genuinely $N$-partite entangled.
It is called fully separable, iff it is $N$-separable.

The concurrence of a pure state $|\psi\rangle \in H=H_{1} \otimes H_{2} \otimes \ldots \otimes H_{N}$ is defined as \cite{PRL92.167902,PRL97.050501}
\begin{equation}
\begin{aligned}
C(|\psi\rangle)=2^{1-N/2}\sqrt{2^{N}-2-\sum\limits_{\alpha}\textrm{Tr}\rho_{\alpha}^{2}},
\end{aligned}
\end{equation}
where the multi-index $\alpha$ runs over all $(2^{N}-2)$ subsets of the $N$ subsystems, and $\rho_{\alpha}$ are the reduced density matrices of all $1$ to $N-1$ partite subsystems. $C$ vanishes exactly for $N$-separable states. The concurrence is extended to mixed state $\rho$ by the so-called convex roof construction,
\begin{equation}
\begin{aligned}
C(\rho) \equiv  \mathop{\textrm{min}}\limits_{\{p_{i},|\psi_{i}\rangle\}} \sum\limits_{i} p_{i}C(|\psi_{i}\rangle),
\end{aligned}
\end{equation}
where the minimization is meant as an optimization over all possible ensemble realizations $\rho=\sum\limits_{i}p_{i}|\psi_{i}\rangle\langle\psi_{i}|$, $p_i\geq 0$ and $\sum\limits_{i}p_{i}=1$. The decomposition attaining the minimum value is said to be the optimal decomposition.

The concurrence \cite{PRL80.2245} of a two-qubit mixed state $\rho$ is given by
\begin{equation}
\begin{aligned}
C(\rho) = \textrm{max}\{\lambda_{1}-\lambda_{2}-\lambda_{3}-\lambda_{4}, 0\}
\end{aligned}
\end{equation}
with the numbers $\lambda_{i}~(i=1, 2, 3, 4)$ are the square roots of the eigenvalues of the non-Hermitian matrix $\rho(\sigma_y\otimes\sigma_y)\rho^*(\sigma_y\otimes\sigma_y)$ in nonincreasing order, where $*$ denotes complex conjugation in the standard basis and $\sigma_y$ is the Pauli matrix.

For $N$-qubit quantum states, a monogamy inequality in terms of squared concurrence was derived in \cite{PRL96.220503}
\begin{equation}
\begin{aligned}
C_{A_{1}|A_{2}A_{3}\ldots A_{N}}^{2}(\rho)\geq \sum\limits_{i=2}^{N}C_{A_{1}A_{i}}^{2}(\rho),
\end{aligned}
\end{equation}
where $C_{A_{1}|A_{2}A_{3}\ldots A_{N}}^{2}(\rho)$ quantifies bipartite entanglement in
the bipartition $A_{1}|A_{2}A_{3}\ldots A_{N}$ and $C_{A_{1}A_{i}}^{2}(\rho)$ characterizes two-qubit entanglement of the reduced state
$\rho_{A_{1}A_{i}}=\textrm{Tr}_{A_{2}\ldots A_{i-1}A_{i+1}\ldots A_{N}}(\rho)$, $i=2,\ldots, N$.

\section{The lower bound of concurrence for four-qubit systems}

We start to discuss the case of four-qubit systems and set up the following theorem. Some explicit examples indicate that the lower bound given by this theorem can be able to improve the known bounds based on detailed comparisons with Theorem 1 in \cite{QIP13.815}.

\emph{Theorem 1.}~~~ For any 4-qubit mixed quantum state $\rho$, the concurrence $C(\rho)$ satisfies
\begin{equation}
\begin{aligned}
C^{2}(\rho) \geq \frac{7}{8} \sum\limits_{i=1}^{3}\sum\limits_{j>i}^{4}C_{ij}^{2}(\rho).
\end{aligned}
\end{equation}

\emph{Proof}.  We begin with pure state. The concurrence of a four-qubit pure state $|\psi\rangle$ can be equivalently written as
\begin{equation}
\begin{aligned}
C^{2}(|\psi\rangle) =\frac{C_{1|234}^{2}+C_{2|134}^{2}+C_{3|124}^{2}+C_{4|123}^{2}+C_{12|34}^{2}+C_{13|24}^{2}+C_{14|23}^{2}}{4}.
\end{aligned}
\end{equation}

By inequality \cite{JMP48.012108,JMP51.112201}
\begin{equation}
\begin{aligned}
C_{12|34}^{2}+C_{13|24}^{2}+C_{14|23}^{2} \geq \frac{3}{4}(C_{1|234}^{2}+C_{2|134}^{2}+C_{3|124}^{2}+C_{4|123}^{2}),
\end{aligned}
\end{equation}
we have
\begin{equation}
\begin{aligned}
C^{2}(|\psi\rangle) &\geq \frac{7}{16}(C_{1|234}^{2}+C_{2|134}^{2}+C_{3|124}^{2}+C_{4|123}^{2})\\
 &\geq \frac{7}{8}(C_{12}^{2}+C_{13}^{2}+C_{14}^{2}+C_{23}^{2}+C_{24}^{2}+C_{34}^{2}),
\end{aligned}
\end{equation}
where the monogamy inequality has been used in the second inequality. Thus, (5) holds for pure state $|\psi\rangle$.

Let $\rho = \sum\limits_{k}p_{k}|\psi_{k}\rangle\langle\psi_{k}|$ be an optimal pure state decomposition for $\rho$ to achieve the minimum value of (2). We have
\begin{equation}
\begin{aligned}
C^{2}(\rho) &= \left\{\sum\limits_{k} p_{k}C(|\psi_{k}\rangle)\right\}^{2}\\
              &\geq \frac{7}{8}\left\{\sum\limits_{k} p_{k}\sqrt{\sum\limits_{1\leqslant i\leqslant3}\sum\limits_{i< j\leqslant4}C_{ij}^{2}(|\psi_{k}\rangle)}\right\}^{2}\\
              &\geq \frac{7}{8}\sum\limits_{1\leqslant i\leqslant3}\sum\limits_{i< j\leqslant4}\left(\sum\limits_{k}p_{k}C_{ij}(|\psi_{k}\rangle)\right)^{2}\\
              &\geq \frac{7}{8}\sum\limits_{1\leqslant i\leqslant3}\sum\limits_{i< j\leqslant4} C_{ij}^{2}(\rho),
\end{aligned}
\end{equation}
where the Minkowski inequality $(\sum\limits_{j}(\sum\limits_{i}x_{ij})^{2})^{\frac{1}{2}} \leq \sum\limits_{i}(\sum\limits_{j}x_{ij}^{2})^{\frac{1}{2}}$ and convexity of concurrence have been used in the second inequality and the third inequality, respectively. The proof is complete.

\emph{Example 1.}~~~~Consider the family of four-qubit states,
$$\rho=\frac{1-t}{16}I_{16}+t|W_{4}\rangle\langle W_{4}|,$$
which is a mixture of four-qubit W state and white noise. Here, $|\text{W}_{4}\rangle = (|0001\rangle+|0010\rangle+|0100\rangle+|1000\rangle)/2$ and $I_{16}$ is the $16\times 16$ identity matrix. By using the formula of concurrence for two-qubit states (3), we get

$$C_{12}(\rho)=C_{13}(\rho)=C_{14}(\rho)=C_{23}(\rho)=C_{24}(\rho)=C_{34}(\rho)=\textrm{max}\left\{0,\frac{t-\sqrt{1-t^{2}}}{2}\right\}.$$

From  Theorem 1, the lower bound of concurrence for this quantum state is given by
$$C^{2}(\rho)\geq \frac{21}{4}C_{12}^{2}(\rho).$$
The lower bound for this quantum state presented in \cite{QIP13.815} is $3C_{12}^{2}(\rho)$. Hence, our bound is better than the one gives in \cite{QIP13.815}.

\emph{Example 2.}~~~~Let us consider the four-qubit state which is the mixture of the Dicke state with two excitations and white noise,
$$\rho=\frac{1-t}{16}I_{16}+t|D_{4}^{2}\rangle\langle D_{4}^{2}|.$$
Here, $|\text{D}_{4}^{2}\rangle = (|0011\rangle+|0101\rangle+|0110\rangle+|1001\rangle+|1010\rangle+|1100\rangle)/\sqrt{6}$. From (3), we have

$$C_{12}(\rho)=C_{13}(\rho)=C_{14}(\rho)=C_{23}(\rho)=C_{24}(\rho)=C_{34}(\rho)=\textrm{max}\left\{0,\frac{5t-3}{6}\right\}.$$

From the lower bound in \cite{QIP13.815}, one has $C^{2}(\rho)\geq 3C_{12}^{2}(\rho)$, while the lower bound of concurrence by Theorem 1 is $$C^{2}(\rho)\geq \frac{21}{4}C_{12}^{2}(\rho).$$  Note that from the lower bound in \cite{PRA85.062320}, $\rho$ is entangled for $t>0.636364$, while by the lower bound  in \cite{QIP13.815},  $\rho$ is entangled for $t>0.618034$.
Therefore, our bound is better than the one in \cite{QIP13.815, PRA85.062320} and detects entanglement better the lower bound \cite{PRA85.062320}.

\emph{Example 3.}~~~~We consider the one-parameter four-qubit state
$$\rho=\frac{1-a}{16}I_{16}+a|\psi\rangle\langle\psi|,$$
where $|\psi\rangle = (|0011\rangle+|0101\rangle+|0110\rangle+|1010\rangle)/2$. It follows from (3),

$$C_{12}(\rho)=C_{14}(\rho)=C_{23}(\rho)=C_{34}(\rho)=\textrm{max}\left\{\frac{a-\sqrt{1-a}}{2},0\right\}$$
and $C_{13}(\rho)=C_{24}(\rho)=0$. From  Theorem 1, the lower bound of concurrence gives
$$C^{2}(\rho)\geq \frac{7}{2}C_{12}^{2}(\rho).$$
From the lower bound in \cite{QIP13.815}, one has $C^{2}(\rho)\geq 2C_{12}^{2}(\rho)$. It shows that our bound is better than that in \cite{QIP13.815}.

\emph{Example 4.}~~~~Consider the four-qubit mixed state $\rho = \frac{1-t}{16}I_{16}+t|\psi\rangle\langle\psi|$, where $|\psi\rangle = (|0000\rangle+|0011\rangle+|1100\rangle+|1111\rangle)/2$. We have
$$C_{12}(\rho)=C_{34}(\rho)=\textrm{max}\left\{0,\frac{3t-1}{2}\right\}$$
and $C_{13}=C_{14}=C_{23}=C_{24}=0$. By our Theorem 1, the lower bound of concurrence is
$$C^{2}(\rho)\geq \frac{7}{4}C_{12}^{2}(\rho),$$
while the lower bound mentioned in \cite{QIP13.815} is $C^{2}(\rho)\geq C_{12}^{2}(\rho)$.
This lower bound can detect entanglement of $\rho$ when $t>\frac{1}{3}$.
Note that, when $t=1$, $C(|\psi\rangle)=\frac{\sqrt{7}}{2}$, namely, the state $|\psi\rangle$ saturates the lower bound. Therefore, our bound is better than the one in \cite{QIP13.815}.

\emph{Note.} The definition of concurrence in \cite{QIP13.815} is different from (1) up to a constant factor $2^{1-N/2}$. In above examples, the difference of the constant factor in defining the concurrence for pure states has already been taken into account.

\section{The lower bounds of concurrence for arbitrary qubit systems}
In this section, we generalize the result of $4$-qubit to any $N$-qubit systems. In the following theorems, a lower bound for arbitrary qubit states is first presented, and then for even-qubit states, a better lower bound is provided with the aid of a monogamy equality.

\emph{Theorem 2.}~~~ For any $N$-qubit ($N \geq 5$) mixed state $\rho$, the concurrence $C(\rho)$ satisfies
\begin{equation}
\begin{aligned}
C^{2}(\rho) \geq \frac{N}{2^{N-2}}\sum\limits_{i=1}^{N-1}\sum\limits_{j>i}^{N}C_{ij}^{2}(\rho).
\end{aligned}
\end{equation}

\emph{Proof.} The concurrence of an $N$-qubit pure state $|\psi\rangle$ can be equivalently written as
\begin{equation}
\begin{aligned}
C^{2}(|\psi\rangle) = \frac{1}{2^{N-1}}\left(\sum\limits_{\{1\}}C_{\{1\}|\{N-1\}}^{2}(|\psi\rangle)+\sum\limits_{\{2\}}C_{\{2\}|\{N-2\}}^{2}(|\psi\rangle)+\ldots+\sum\limits_{\{N-1\}}C_{\{N-1\}|\{1\}}^{2}(|\psi\rangle)\right),
\end{aligned}
\end{equation}
where $C_{\{j\}|\{N-j\}}$ is the concurrence of the state obtained by tracing out $j$ qubits from the state $|\psi\rangle$
and $\sum\limits_{\{j\}}$ is taken over all combinations $\{j\}$ of $j$ indices.

From inequality \cite{JMP48.012108}
\begin{equation}
\begin{aligned}
\sum\limits_{\{2\}}C_{\{2\}|\{N-2\}}^{2}(|\psi\rangle)\geq \frac{N-2}{2}\sum\limits_{\{1\}}C_{\{1\}|\{N-1\}}^{2}(|\psi\rangle),
\end{aligned}
\end{equation}
we get
\begin{equation}
\begin{aligned}
C^{2}(|\psi\rangle) \geq  \frac{N}{2^{N-1}}\sum\limits_{\{1\}}C_{\{1\}|\{N-1\}}^{2}(|\psi\rangle) \geq \frac{N}{2^{N-2}}\sum\limits_{i=1}^{N-1}\sum\limits_{j>i}^{N}C_{ij}^{2}(|\psi\rangle).
\end{aligned}
\end{equation}

Similar to the proof of Theorem 1, (10) is also right for any $N$-qubit ($N \geq 5$) mixed state $\rho$.

In \cite{PRL114.140402}, a monogamy equality for qubit entanglement has been proven, we get an improved lower bound of concurrence for even-qubit by using this equality.

\emph{Theorem 3.}~~~ For any even-qubit ($N \geq6$) mixed state $\rho$, the concurrence $C(\rho)$ satisfies
\begin{equation}
\begin{aligned}
C^{2}(\rho) \geq \frac{N-2}{2^{N-3}}\sum\limits_{i=1}^{N-1}\sum\limits_{j>i}^{N}C_{ij}^{2}(\rho).
\end{aligned}
\end{equation}

\emph{Proof.} For any $N$-qubit pure state, an exact monogamy relation was presented in \cite{PRL114.140402},

\begin{equation}
\begin{aligned}
2|H(|\psi\rangle)|^{2} = \sum\limits_{\{1\}}C_{\{1\}|\{N-1\}}^{2}(|\psi\rangle) - \sum\limits_{\{2\}}C_{\{2\}|\{N-2\}}^{2}(|\psi\rangle) +\cdots +(-1)^{N}\sum\limits_{\{N-1\}}C_{\{N-1\}|\{1\}}^{2}(|\psi\rangle),
\end{aligned}
\end{equation}
where $H(|\psi\rangle)$ is a well known polynomial invariant, and defined by $H(|\psi\rangle)=\langle\psi|\tilde{\psi\rangle}$, $|\tilde{\psi\rangle} = \sigma_{y}^{\otimes n}|\psi^{*}\rangle$ \cite{PRA63.044301}.

For even-qubit pure state, one has
\begin{equation}
\begin{aligned}
C^{2}(|\psi\rangle) = \frac{1}{2^{N-2}}\left\{\sum\limits_{\{1\}}C_{\{1\}|\{N-1\}}^{2}(|\psi\rangle) + \sum\limits_{\{2\}}C_{\{2\}|\{N-2\}}^{2}(|\psi\rangle) +\cdots +\frac{1}{2}\sum\limits_{\{\frac{N}{2}\}}C_{\{\frac{N}{2}\}|\{\frac{N}{2}\}}^{2}(|\psi\rangle)\right\}.
\end{aligned}
\end{equation}

From (12), (15) and (16), we obtain
\begin{equation}
\begin{aligned}
C^{2}(|\psi\rangle) \geq \frac{2}{2^{N-2}}\sum\limits_{\{2\}}C_{\{2\}|\{N-2\}}^{2}(|\psi\rangle) \geq \frac{N-2}{2^{N-2}}\sum\limits_{\{1\}}C_{\{1\}|\{N-1\}}^{2}(|\psi\rangle) \geq \frac{N-2}{2^{N-3}}\sum\limits_{i=1}^{N-1}\sum\limits_{j>i}^{N}C_{ij}^{2}(|\psi\rangle).
\end{aligned}
\end{equation}

Similar to the proof of Theorem 1, (14) holds for any even-qubit ($N \geq 6$) mixed state $\rho$.

\section{The detection of $k$-nonseparability of multipartite quantum systems}
It is widely known that multipartite concurrence is zero for fully separable states, i.e., multipartite state is entangled (not fully separable) if concurrence is greater than zero. In this section, we show that multipartite concurrence can also play a useful role in the $k$-nonseparability problem based on the norms of the correlation tensors in the generalized Bloch representation of an $N$-partite quantum state.

\emph{Theorem 4.}~~~ Any $N$ qudit state $\rho$ is $k$-nonseparable if the concurrence $C(\rho)$ satisfies
\begin{equation}
C(\rho)>
\begin{cases}
   2^{1-N/2}\sqrt{2^{N}-2^{k}+\frac{2^{k}-2}{d}-2\sum\limits_{i=1}^{(N-1)/2}\frac{C_{N}^{i}}{d^{i}}} & \text{for odd $N$}, \\
   2^{1-N/2}\sqrt{2^{N}-2^{k}+\frac{2^{k}-2}{d}-2\sum\limits_{i=1}^{N/2-1}\frac{C_{N}^{i}}{d^{i}}-\frac{C_{N}^{N/2}}{d^{N/2}}} & \text{for even $N$},
 \end{cases}
\end{equation}
where $C_{N}^{i}=N!/[i!(N-i)!]$.

\emph{Proof.} An $N$-partite state $\rho$ acting on $H_{1} \otimes H_{2} \otimes \ldots \otimes H_{N}~(\text{dim}H_{i}=d)$ can be written as \cite{QIC08.41}
\begin{equation*}
\begin{aligned}
\rho=&\frac{1}{d^{N}}\Big(\bigotimes_{j=1}^{N}I_{d}+\sum\limits_{\{\mu_{1}\}}\sum\limits_{\alpha_{1}}\mathcal{T}_{\alpha_{1}}^{\{\mu_{1}\}}\lambda_{\alpha_{1}}^{\{\mu_{1}\}}+\sum\limits_{\{\mu_{1}\mu_{2}\}}\sum\limits_{\alpha_{1}\alpha_{2}}\mathcal{T}_{\alpha_{1}\alpha_{2}}^{\{\mu_{1}\mu_{2}\}}\lambda_{\alpha_{1}}^{\{\mu_{1}\}}\lambda_{\alpha_{2}}^{\{\mu_{2}\}}
+\sum\limits_{\{\mu_{1}\mu_{2}\mu_{3}\}}\sum\limits_{\alpha_{1}\alpha_{2}\alpha_{3}}\mathcal{T}_{\alpha_{1}\alpha_{2}\alpha_{3}}^{\{\mu_{1}\mu_{2}\mu_{3}\}}\lambda_{\alpha_{1}}^{\{\mu_{1}\}}\lambda_{\alpha_{2}}^{\{\mu_{2}\}}\lambda_{\alpha_{3}}^{\{\mu_{3}\}}\\
     &+\cdots+\sum\limits_{\{\mu_{1}\mu_{2}\cdots\mu_{M}\}}\sum\limits_{\alpha_{1}\alpha_{2}\cdots\alpha_{M}}\mathcal{T}_{\alpha_{1}\alpha_{2}\cdots\alpha_{M}}^{\{\mu_{1}\mu_{2}\cdots\mu_{M}\}}\lambda_{\alpha_{1}}^{\{\mu_{1}\}}\lambda_{\alpha_{2}}^{\{\mu_{2}\}}\cdots\lambda_{\alpha_{M}}^{\{\mu_{M}\}}
+\cdots+\sum_{\alpha_{1}\alpha_{2}\cdots\alpha_{N}}\mathcal{T}_{\alpha_{1}\alpha_{2}\cdots\alpha_{N}}^{\{1,2,\cdots,N\}}\lambda_{\alpha_{1}}^{\{1\}}\lambda_{\alpha_{2}}^{\{2\}}\cdots\lambda_{\alpha_{N}}^{\{N\}}\Big),
\end{aligned}
\end{equation*}
where $\lambda_{\alpha_{l}}$ are the $SU(d)$ generators, $\{\mu_{1}\mu_{2}\cdots\mu_{M}\}$ is a subset of $\{1,2,\cdots,N\}$, $\lambda_{\alpha_{l}}^{\{\mu_{l}\}}=I_{d}\otimes \cdots \otimes I_{d} \otimes \lambda_{\alpha_{l}}\otimes I_{d}\otimes \cdots \otimes I_{d}$ with $\lambda_{\alpha_{l}}$ appearing at the $\mu_{l}$th position, $I_{d}$ is the $d\times d$ identity matrix, and
$$\mathcal{T}_{\alpha_{1}\alpha_{2}\cdots\alpha_{M}}^{\{\mu_{1}\mu_{2}\cdots\mu_{M}\}}=\frac{d^{M}}{2^{M}}\textrm{Tr}[\rho\lambda_{\alpha_{1}}^{\{\mu_{1}\}}\lambda_{\alpha_{2}}^{\{\mu_{2}\}}\cdots\lambda_{\alpha_{M}}^{\{\mu_{M}\}}],$$
which can be viewed as the entries of tensors $\mathcal{T}^{\{\mu_{1}\mu_{2}\cdots\mu_{M}\}}$.

We first discuss an $N$-partite pure state $|\psi\rangle$.
Assume that $|\psi\rangle \in H_{1} \otimes H_{2} \otimes \ldots \otimes H_{N}$ is $k$-separable with regard to $k$-partition $A_{1}|A_{2}|\cdots |A_{k}$. Let $||\cdot||$ denote the Euclidean norm for a tensor, $|A_{i}|$  the number of elements of $A_{i}$, and $|A|$  the minimum value of $|A_{i}|$. Then  $\textrm{Tr}(\rho_{A_{1}}^{2})=\textrm{Tr}(\rho_{A_{2}}^{2})=\cdots=\textrm{Tr}(\rho_{A_{k}}^{2})=1$, i.e.,
\begin{equation*}
\begin{aligned}
\frac{2}{d^{|A_{i}|}+1}\sum_{j\in \{1,2,\ldots,|A_{i}|\}}||\mathcal{T}^{i_{j}}||^{2}+\frac{2^{2}}{d^{|A_{i}|}+2}\sum_{j,l}||\mathcal{T}^{i_{j}i_{l}}||^{2}+\cdots+\frac{2^{|A_{i}|}}{d^{2|A_{i}|}}||\mathcal{T}^{i_{1}\cdots i_{|A_{i}|}}||^{2}=1-\frac{1}{d^{|A_{i}|}}, \end{aligned}
\end{equation*}
where $i\in \{1,2,\ldots,k\}$. Thus we derive in a similar way to \cite{PRA92.062338}  that  for odd $N$,
\begin{equation*}
\begin{aligned}
\sum_{\alpha=1}^{2^{N}-2}\textrm{Tr}\rho_{\alpha}^{2}
&\geqslant 2\Big[C_{N}^{1}\frac{1}{d}+C_{N}^{2}\frac{1}{d^{2}}+\cdot+C_{N}^{\frac{N-1}{2}}\frac{1}{d^{(N-1)/2}}\Big]+(C_{k}^{1}+C_{k}^{2}+\cdots +C_{k}^{k-1})\Big(1-\frac{1}{d^{|A|}}\Big),\\
&= 2\Big[C_{N}^{1}\frac{1}{d}+C_{N}^{2}\frac{1}{d^{2}}+\cdot+C_{N}^{\frac{N-1}{2}}\frac{1}{d^{(N-1)/2}}\Big]+(2^{k}-2)\Big(1-\frac{1}{d^{|A|}}\Big),
\end{aligned}
\end{equation*}
while for even $N$,
\begin{equation*}
\begin{aligned}
\sum_{\alpha=1}^{2^{N}-2}\textrm{Tr}\rho_{\alpha}^{2}\geqslant
2\Big[C_{N}^{1}\frac{1}{d}+C_{N}^{2}\frac{1}{d^{2}}+\cdot+C_{N}^{\frac{N-2}{2}}\frac{1}{d^{(N-2)/2}}\Big]+C_{N}^{\frac{N}{2}}\frac{1}{d^{N/2}}+(2^{k}-2)\Big(1-\frac{1}{d^{|A|}}\Big).
\end{aligned}
\end{equation*}

Therefore, for $k$-separable pure state, we obtain

\begin{equation}
\begin{aligned}
C(|\psi\rangle)&=2^{1-N/2}\sqrt{2^{N}-2-\sum\limits_{\alpha}\textrm{Tr}\rho_{\alpha}^{2}}\\
&\leqslant \begin{cases}
   2^{1-N/2}\sqrt{2^{N}-2^{k}+\frac{2^{k}-2}{d^{|A|}}-2\sum\limits_{i=1}^{(N-1)/2}\frac{C_{N}^{i}}{d^{i}}} & \text{for odd $N$}, \\
   2^{1-N/2}\sqrt{2^{N}-2^{k}+\frac{2^{k}-2}{d^{|A|}}-2\sum\limits_{i=1}^{N/2-1}\frac{C_{N}^{i}}{d^{i}}-\frac{C_{N}^{N/2}}{d^{N/2}}} & \text{for even $N$.}
 \end{cases}
\end{aligned}
\end{equation}

If $\rho=\sum\limits_{i}p_{i}|\psi_{i}\rangle\langle\psi_{i}|$ is a $k$-separable mixed state, then by using of the convexity of the concurrence $C(\rho)\leqslant \sum\limits_{i}p_{i}C(|\psi_{i}\rangle)$ and inequality (19), we get
\begin{equation}
C(\rho)\leqslant
\begin{cases}
   2^{1-N/2}\sqrt{2^{N}-2^{k}+\frac{2^{k}-2}{d}-2\sum\limits_{i=1}^{(N-1)/2}\frac{C_{N}^{i}}{d^{i}}} & \text{for odd $N$}, \\
   2^{1-N/2}\sqrt{2^{N}-2^{k}+\frac{2^{k}-2}{d}-2\sum\limits_{i=1}^{N/2-1}\frac{C_{N}^{i}}{d^{i}}-\frac{C_{N}^{N/2}}{d^{N/2}}} & \text{for even $N$}.
 \end{cases}
\end{equation}

$\emph{Remark.}$~~~~Theorem obtained in \cite{PRA92.062338} is the special case $k=2$ of Theorem 4.

\emph{Example 5.}~~~~We consider the quantum state in Example 4. When $k=3$, the lower bound in (18) is $\frac{\sqrt{22}}{4}$. By using Theorem 1 and Theorem 4, the $3$-partite nonseparable is detected for $t>0.9243$.

\emph{Example 6.}~~~~Consider the family of $n$-qubit states,
$$\rho=\frac{1-p}{2^{n}}I_{2^{n}}+p|\text{GHZ}_{n}\rangle\langle \text{GHZ}_{n}|,$$
which is a mixture of $n$-qubit GHZ state and white noise. Here, $|\text{GHZ}_{n}\rangle=(|00\cdots 0\rangle+|11\cdots 1\rangle)/\sqrt{2}$.
Applying Observation 1 in \cite{PRL109.200503} and Proposition in \cite{PRA92.062338}, we get
$$C^{2}(\rho)\geq \frac{2^{n-1}-1}{2^{n-2}}\left[\frac{(2^{n-1}+1)p-1}{2^{n-1}}\right]^{2}.$$
From this bound, the state $\rho$ is entangled (not fully separable) for $p>1/(2^{n-1}+1)$. For $p=1$, this reproduces exactly concurrence of the pure $n$-qubit GHZ state. This means that we provide a necessary and sufficient criterion
for entanglement for the family of states $\rho$, since it is known that these states are fully separable iff $0\leq p\leq 1/(2^{n-1}+1)$~\cite{PRA61.042314,JMO47.387}. Because concurrence of mixed state is a convex function and this bound coincides with the
exact value on the point $p=1/(2^{n-1}+1)$ and $p=1$, so the bound equals the exact value on the whole interval $p\in [1/(2^{n-1}+1),1]$. Namely, for the states of $n$-qubit GHZ state mixed with white noise, we present exact value of concurrence
\begin{equation}\label{}
   C(\rho)= \sqrt{\frac{2^{n-1}-1}{2^{n-2}}}\frac{(2^{n-1}+1)p-1}{2^{n-1}}, ~~ p\in \Big[\frac{1}{2^{n-1}+1},1\Big].
\end{equation}
Applying Theorem 4, we can detect $k$-nonseparability of $n$-qubit GHZ state mixed with white noise. For instance, when $n=4$ and $k=3$, the $3$-partite nonseparable is detected for $p>0.8991$.

\section{Conclusion}

In summary we have derived a new lower bound of concurrence for four-qubit states based on the monogamy inequality of entanglement. Through specific examples, we have illustrated that the new lower bound may be used to improve the previous lower bounds. We have also seen how to generalize this approach to $N$-qubit states to obtain the lower bound of the concurrence. Moreover, we have obtained an exact formula of concurrence for some mixed quantum states. For even-qubit states,  an improved lower bound of concurrence using a monogamy equality was derived. Compared with the lower bound of the concurrence presented in \cite{QIP13.815}, our advantage is easy to compute without considering the choice of parameters. At the same time, we prove that a multipartite state is $k$-nonseparable if the multipartite concurrence is larger than a constant related to the value of $k$,  the qudit number and the dimension of the subsystems. Our result may support a new way to detect the multipartite $k$-nonseparable states.

\acknowledgments
This work was supported by the National Natural Science Foundation
of China under Grant Nos: 11371005, 11475054, Hebei Natural Science Foundation
of China under Grant Nos: A2014205060, A2016205145.

\end{document}